\documentstyle[12pt]{article}
\newcommand{\be}{\begin{eqnarray}}
\newcommand{\ee}{\end{eqnarray}}

\input{epsfig.sty}

\begin{document}

\Huge{\noindent{Istituto\\Nazionale\\Fisica\\Nucleare}}

\vspace{-3.9cm}

\Large{\rightline{Sezione SANIT\`{A}}}
\normalsize{}
\rightline{Istituto Superiore di Sanit\`{a}}
\rightline{Viale Regina Elena 299}
\rightline{I-00161 Roma, Italy}

\vspace{0.65cm}

\rightline{INFN-ISS 96/1}
\rightline{January 1996}

\vspace{2cm}

\begin{center}

\LARGE{Calculation of the Isgur-Wise function from a light-front
constituent quark model \footnote{{\normalsize \bf ~ To appear
in Physics Letters B (1996).}}}\\

\vspace{1cm}

\large{S. Simula}\\

\vspace{0.5cm}

\normalsize{Istituto Nazionale di Fisica Nucleare, Sezione
Sanit\`{a}\\ Viale Regina Elena 299, I-00161 Roma, Italy}

\end{center}

\vspace{1cm}

\begin{abstract}

The space-like elastic form factor of heavy-light pseudoscalar
mesons is investigated within a light-front constituent quark model
in order to evaluate the Isgur-Wise form factor. The relativistic
composition of the constituent quark spins is properly taken into
account using the Melosh rotations, and various heavy-meson wave
function are considered, including the eigenfunctions of an
effective light-front mass operator reproducing meson mass spectra.
It is shown that in a wide range of values of the recoil the
Isgur-Wise form factor exhibits a moderate dependence upon the
choice of the heavy-meson wave function and is mainly governed by
the effects of the confinement scale.  

\end{abstract}

\vspace{1cm}


\newpage

\pagestyle{plain}

\indent Weak decays of hadrons can provide relevant information on
the fundamental parameters of the standard model of the
electroweak interaction and on the internal structure of hadrons.
The extraction of the Cabibbo-Kobayashi-Maskawa \cite{CKM} matrix
elements from the experiments requires therefore a precise
knowledge of electroweak hadron form factors. As is well known, in
case of hadrons containing a single heavy-quark $Q$ ($m_Q \gg
\Lambda_{QCD}$) the complexity of the theoretical analysis is
strongly reduced by the Heavy Quark Symmetry ($HQS$), i.e. a
spin-flavour symmetry that is a consequence of $QCD$ in the limit
of infinite quark masses \cite{IW89,HQET}. The $HQS$ requires that,
when $m_Q \rightarrow \infty$, all the non-perturbative strong
physics describing the weak decays of heavy hadrons is contained in
a single universal function, known as the Isgur-Wise ($IW$) form
factor $\xi^{(IW)}(\omega)$, where $\omega \equiv v \cdot v'$ and
$v_{\mu}$ ($v'_{\mu}$) is the four-velocity of the initial (final)
hadron. However, the $HQS$ does not help in predicting the $IW$
function itself, so that several attempts  \cite{NEU94} have been
made to calculate $\xi^{(IW)}(\omega)$ both from the fundamental 
theory and models, like the quark models of Refs.
\cite{ISGW,BSW,CW94}. In such models a simple gaussian-like ansatz
for the heavy-meson wave function has been adopted. Furthermore,
the relativistic treatment of the light-quark spin requires a
particular care; as a matter of fact, it has been shown \cite{CW94}
that relativistic effects remarkably increase the slope of the $IW$
form factor at the zero-recoil point ($\rho^2 \equiv -
[d\xi^{(IW)}(\omega) / d\omega]_{\omega = 1}$).

\indent The aim of this letter is to investigate the $IW$ form
factor within a light-front constituent quark ($CQ$) model, where:
i) the relativistic composition of the $CQ$ spins is properly taken
into account using the Melosh rotations \cite{MEL74}, and ii)
various heavy-meson wave function are considered, including the
eigenfunctions of a light-front mass operator, constructed from the
effective $q \bar{q}$ interaction of Godfrey and Isgur ($GI$)
\cite{GI85}, which nicely fits meson mass spectra. Our $CQ$ model
has been already applied to the investigation of the leptonic decay
constant of light and heavy pseudoscalar ($PS$) mesons
\cite{CAR94}, the electromagnetic (e.m.) form factors of light $PS$
\cite{CAR94} and vector \cite{CAR95} mesons, the nucleon elastic
and $N - \Delta$ transition e.m. form factors \cite{CPSS95}. In this
letter, the $CQ$ model formulated on the light-front is used to
investigate the elastic form factor of heavy-light $PS$ mesons at
space-like values of the squared four-momentum transfer $q^2 \equiv
q \cdot q \leq 0$ in order to evaluate the $IW$ form factor. The
choice of the space-like sector is motivated by the fact that, only
for $q^2 \leq 0$, the contribution of the so-called Z-graph (pair
creation from the vacuum) can be suppressed by choosing an
appropriate reference frame  \cite{ZGRAPH} (viz. a frame in which
$q^+ = q^0 + \hat{n} \cdot \vec{q} = 0$ where the vector $\hat{n} =
(0,0,1)$ defines the spin quantization axis). In a wide range of
values of the recoil the $IW$ form factor is found to be mainly
governed by the effects of the confinement scale and, in
particular, it is slightly affected by the high-momentum components
generated in the heavy-meson wave function by the
one-gluon-exchange ($OGE$) term of the effective $GI$ interaction. 

\indent As is well known (cf. \cite{LIGHT-FRONT}), light-front
hadron wave functions are eigenfunctions of the mass operator
${\cal{M}} = M_0 + {\cal{V}}$ and of the non-interacting angular
momentum operators $j^2$ and $j_n$, where $M_0$ is the free-mass
operator and ${\cal{V}}$ a Poincar\'e-invariant interaction term.
The operator $M_0$ reads as $M_0^2 = {k_{\perp}^2 + {m_q}^2 \over
\xi} + {k_{\perp}^2 + {m_{\bar{q}}}^2 \over 1 - \xi }$, where
$m_q$($m_{\bar{q}}$) is the constituent quark (antiquark) mass and
the intrinsic light-front variables are $\vec{k}_{\perp} =
\vec{p}_{q \perp} - \xi \vec{P}_{\perp}$ and $\xi = p^+_q / P^+$,
where the subscript $\perp$ indicates the projection perpendicular
to $\hat{n}$ and the {\em plus} component of a 4-vector $p \equiv
(p^0, \vec{p})$ is given by $p^+ = p^0 + \hat{n} \cdot \vec{p}$.
Finally, $\tilde{P} \equiv (P^+, \vec{P}_{\perp}) = \tilde{p}_q +
\tilde{p}_{\bar{q}}$ is the light-front meson momentum and
$\tilde{p}_q$ the $CQ$ one. It should be pointed out that the
centre-of-mass motion is exactly factorized out. Moreover, in terms
of the longitudinal momentum $k_n$, defined as $k_n = (\xi - 1/2)
M_0 + (m_{\bar{q}}^2 - m_q^2) / 2M_0$, the free-mass $M_0$ acquires
a more familiar form, viz. $M_0 = \sqrt{m_q^2 + k^2 } +
\sqrt{m_{\bar{q}}^2 + k^2 }$, with $k^2 = k_{\perp}^2 + k_n^2$.
Omitting for the sake of simplicity the colour degrees of freedom,
the requirement of Poincar\`{e} covariance for the intrinsic wave
function $\chi (\xi, \vec{k}_{\perp}, \nu \bar{\nu})$ of a $PS$
meson implies (cf., e.g., \cite{CAR94})
 \be
    \chi(\vec{k}_{\perp}, \xi,\nu \bar{\nu}) =
    {\cal{R}}(\vec{k}_{\perp}, \xi, \nu \bar{\nu}) ~ w^{PS}(k^2) ~
    \sqrt{J(\vec{k}_{\perp}, \xi)}
    \label{WF}
 \ee
where $\nu, \bar{\nu}$ are the $CQ$ spin variables and
$J(\vec{k}_{\perp}, \xi) =$ ${M_0 \over 16\pi \xi (1 - \xi)} \left [
1 - \left ( m^2_q - m^2_{\bar{q}} \over M^2_0 \right ) ^2 \right ]$
is the Jacobian of the transformation between $k_n$ and $\xi$. In
Eq. (\ref{WF}) the momentum-dependent quantity ${\cal{R}}$ is a
unitary matrix arising from the Melosh rotations of the $CQ$
spins; its explicit expression can be found, e.g., in \cite{JAUS}.

\indent The wave function $w^{PS}(k^2) ~ |0 0 \rangle$,
where $| 0 0 \rangle = \sum_{\nu \bar{\nu}} ~ \langle {1 \over 2}
\nu {1 \over 2} \bar{\nu} |0 0 \rangle |{1 \over 2} \nu \rangle |
{1 \over 2} \bar{\nu} \rangle$ is the canonical quark-spin wave
function, is eigenfunction of the transformed mass operator $M =
{\cal{R}} {\cal{MR}}^{\dag} = {\cal{R}} M_0 {\cal{R}}^{\dag} +
{\cal{R}} {\cal{VR}}^{\dag}$. Since the free-mass commutes with the
Melosh rotation, one has $M = M_0 + V$, where the interaction $V =
{\cal{R}} {\cal{VR}}^{\dag}$ has to be invariant upon rotations and
independent of the total momentum and the centre-of-mass coordinates
(cf. \cite{LIGHT-FRONT}). Following Refs. \cite{CAR94,CAR95}, the
Melosh-rotated mass operator $M$ is chosen to be the effective
Hamiltonian, $H_{q \bar{q}}$, proposed in \cite{GI85} for the
description of meson mass spectra. Therefore, the radial wave
function $w^{PS}(k^2)$, appearing in Eq. (\ref{WF}), is
eigenfunction of the effective $GI$ Hamiltonian, viz.
 \be
    H_{q \bar{q}} ~ w^{PS}(k^2) | 0 0 \rangle \equiv \left
    [\sqrt{m_q^2 + k^2} + \sqrt{m_{\bar{q}}^2 + k^2} + V_{q \bar{q}}
    \right ] ~ w^{PS}(k^2) | 0 0 \rangle  =  M_{PS} ~ w^{PS}(k^2) |
    0 0 \rangle
    \label{Hqq}
 \ee
where $M_{PS}$ is the mass of the $PS$ meson and $V_{q \bar{q}}$ the
$GI$ effective $q \bar{q}$  potential, composed by a $OGE$ term
(dominant at short separations) and a linear-confining term
(dominant at large separations). In what follows, three different
forms of $w^{PS}$ will be considered and labeled as $w^{PS}_{(GI)}$,
$w^{PS}_{(conf)}$ and $w^{PS}_{(HO)}$. The first two correspond to
the solutions of Eq. (\ref{Hqq}) obtained using for $V_{q \bar{q}}$
the full $GI$ interaction and only its linear-confining term,
respectively. The third one is a simple gaussian-like ansatz
$w^{PS}_{(HO)} \propto e^{-k^2 / (2 \beta_{PS}^2)}$, where the
harmonic oscillator ($HO$) parameter $\beta_{PS}$ has been fixed by
requiring that the average value of $k_{\perp}^2$ calculated with
$w^{PS}_{(conf)}$ and $w^{PS}_{(HO)}$ is the same. In this way
$w^{PS}_{(conf)}$ and $w^{PS}_{(HO)}$ correspond to a linear- and
quadratic-confining potential with the same scale, respectively. In
case of the $D$ ($D_s$), $B$ ($B_s$) and  $T$ ($T_s$) mesons, the
value of $\beta_{PS}$ ($ = \sqrt{<k_{\perp}^2>_{conf}}$) is:
$\beta_{D(D_s)} = 0.386 ~ (0.408) ~ (GeV/c)$, $\beta_{B(B_s)} =
0.417 ~ (0.445) ~ (GeV/c)$ and $\beta_{T(T_s)} = 0.440 ~ (0.472) ~
(GeV/c)$. For the above-mentioned mesons Eq. (\ref{Hqq}) has been
solved by expanding the wave function onto a (truncated) set of
$HO$ basis states and by applying the Raleigh-Ritz variational
principle to the coefficients of the expansion. It has been checked
that the convergence for all the quantities considered in this
letter is reached completely when all the basis states up to $40$
HO excitation quanta are included in the expansion. The $CQ$'s
masses are fixed at the values adopted in \cite{GI85} to reproduce
meson mass spectra, namely: $ m_u = m_d = 0.220 ~ GeV, ~ m_s =
0.419 ~ GeV, ~ m_c = 1.628 ~ GeV$ and $m_b = 4.977 ~ GeV$. As to
the $t$-quark mass, the value $m_t = 160 ~ GeV$ is considered. It
should be pointed out that the "inertia" parameter $\bar{\Lambda}
\equiv M_{PS} - m_Q$, often introduced in the Heavy Quark Effective
Theory \cite{NEU94}, calculated using the $GI$ interaction for the
$T$ ($T_s$) meson, is found to be $0.38 ~ (0.46) ~ GeV$, in
agreement with the expectations of recent lattice $QCD$ simulations
\cite{DAV94}.  

\indent The $CQ$ momentum distribution $|w^{PS}(k^2)|^2$, calculated
for the $D$, $B$ and $T$ mesons using $w^{PS}_{(conf)}$ and
$w^{PS}_{(GI)}$, is shown in Fig. 1(a), while the comparison among
$|w^B_{(HO)}|^2$, $|w_{(conf)}^B|^2$ and $|w_{(GI)}^B|^2$ is
illustrated in Fig. 1(b), where the gaussian-like wave function of
Ref. \cite{ISGW} is also reported. It can be seen that: i) the
$OGE$ part of the $GI$ effective interaction generates a huge
amount of high-momentum components at $k > 1 ~ (GeV/c)$ (cf.
\cite{CAR94}); ii) the  behaviours of $w_{(HO)}^{PS}$ and
$w_{(conf)}^{PS}$ are very similar at low momenta ($k < 1 ~
(GeV/c)$); iii) the wave functions $w_{(HO)}^{PS}$ do not differ
appreciably from the wave functions adopted in \cite{ISGW}. Similar
results hold as well for the $CQ$ momentum distribution in $D_s$,
$B_s$ and $T_s$ mesons. It should be pointed out that the $OGE$
term of the $GI$ interaction generates high-momentum components
also in the wave function of light mesons and baryons, and sharply
affects their e.m. form factors both at low and high $q^2$  (see
\cite{CAR94,CAR95,CPSS95}).

\indent Within the light-front formalism any e.m. form factor
can be evaluated in terms of the matrix elements of the plus
component of the e.m. current operator \cite{LIGHT-FRONT}. We stress
again that for space-like values of $q^2$ a reference frame where
$q^+ = 0$ can be chosen in order to suppress the contribution of
the pair creation from the vacuum \cite{ZGRAPH}. In what follows the
heavy quark $Q$ is assumed to be point-like and a one-body
approximation for the e.m. current is adopted. Therefore, the $IW$
function is simply related to the matrix element of the "good"
component of the e.m. heavy-quark current, $\bar{Q} ~ \gamma^+ ~ Q$,
by
 \be
    \xi^{(IW)}(\omega) = lim_{m_Q \rightarrow \infty} ~
    H_{PS}^Q(q^2) = lim_{m_Q \rightarrow \infty} ~ \langle P'|
    \bar{Q} ~ \gamma^+ ~ Q |P \rangle / 2P^+
    \label{IW}
\ee
where $q^2 = 2 M_{PS}^2 ~ (1 - \omega)$. The explicit expression for
$H_{PS}^Q(q^2)$ is given by (cf. \cite{CAR94})
 \be
    H_{PS}^Q(q^2) = \int d\vec{k}_{\perp} d\xi ~
    \sqrt{J(\vec{k}_{\perp}, \xi) ~ J(\vec{k'}_{\perp}, \xi)} ~
    w^{PS}(k^2) ~ w^{PS}({k'}^2) ~ R(\vec{k}_{\perp},
    \vec{k'}_{\perp}, \xi)
    \label{BODY}
 \ee
with $\vec{k'}_\perp = \vec{k}_\perp + (1 - \xi) \vec{q}_\perp$, 
${M'}_0^2 = {{k'}_{\perp}^2 + m_Q^2 \over \xi} + {{k'}_{\perp}^2 +
m_{sp}^2 \over (1 - \xi)}$, $k'_n = (\xi - {1 \over 2}) M'_0 +
{m_{sp}^2 - m_Q^2 \over 2 M'_0}$ and $|\vec{q}_{\perp}|^2 = - q^2$.
In Eq. (\ref{BODY}) $R$ is the contribution of the Melosh rotations
and reads as
 \be
   R(\vec{k}_{\perp}, \vec{k'}_{\perp}, \xi) = {\xi (1 - \xi) ~
   [M_0^2 - (m_Q - m_{sp})^2] + \vec{k}_\perp \cdot (\vec{k'}_\perp
   - \vec{k}_{\perp}) \over \xi (1 - \xi) ~ \sqrt{M_0^2 - (m_Q -
   m_{sp})^2} ~ \sqrt{{M'}_0^2 - (m_Q - m_{sp})^2}}
   \label{MELOSH}
 \ee
where $m_{sp}$ is the spectator-quark mass. Since the radial wave
function  is normalized as $\int_0^{\infty} dk k^2 ~ |w^{PS}(k^2)|^2
= 1$, one has $H_{PS}^Q(\omega = 1) = 1$ both at finite and
infinite heavy-quark masses.

\indent The values of the slope of the body form factor $H_{PS}^Q$
at the zero-recoil point, $\rho_Q^2 \equiv - [d H_{PS}^Q (\omega) /
d \omega]_{\omega = 1}$, calculated using $w_{(HO)}^{PS}$,
$w_{(conf)}^{PS}$ and $w_{(GI)}^{PS}$, are reported in Table 1 (2)
for the $D ~ (D_s)$, $B ~ (B_s)$ and $T ~ (T_s)$ mesons. The
results of the extrapolation to infinite $PS$ meson masses,
obtained through a simple quadratic polynomial in terms of $1 /
M_{PS}$, are also shown and compared with the predictions of the
$ISGW$ \cite{ISGW} and the relativistic flux tube ($RFT$)
\cite{RFT} models, as well as with recent $QCD$ sum rule
\cite{BS93,NAR94} and lattice $QCD$ results \cite{LAT1,LAT2}. It
can be seen that: i) in the limit of infinite heavy-quark masses
the choice of the heavy-meson wave function affects the slope only
by $\sim 10 \div 25 \%$; ii) the asymptotic value is reached from
below and, in particular, the slope of $H_{PS = D(D_s)}^{Q = c}$ is
$\sim 10 \div 30 \%$ lower than its asymptotic value; iii) the
effects of the Melosh rotations of the $CQ$ spins (which are
necessary for ensuring the correct transformation properties of the
meson wave function under kinematical light-front boosts) are
relevant both at finite and infinite heavy-quark masses; iv) the
slope of the $IW$ form factor, calculated at $m_{sp} = 0.220 ~
(0.419) ~ GeV$, is $\rho^2 = 1.03 ~ (1.14)$, i.e. it is larger than
the $ISGW$ prediction $\rho^2 \sim 0.6 ~ (0.8)$ \cite{ISGW} and in
agreement with both $QCD$ sum rule results \cite{BS93,NAR94} and
lattice $QCD$ simulations \cite{LAT1,LAT2} as well as with the
prediction of Ref. \cite{RFT}. The $\omega$-dependence of the body
form factor $H_{PS}^Q$, calculated for the $D$, $B$ and $T$ mesons
using $w_{(conf)}^{PS}$ and $w_{(GI)}^{PS}$, is illustrated in Fig.
2. The results obtained with $w_{(HO)}^{PS}$ turn out to be very
similar to the ones calculated using $w_{(conf)}^{PS}$. From Fig. 2
it can be seen that, in a wide range of values of the recoil, the
Melosh composition of the $CQ$ spins remarkably affects the
calculations performed both with and without the high-momentum
components generated in the heavy-meson wave function by the $OGE$
interaction. Moreover, the results obtained at the $T$-meson mass
($\sim 160 ~ GeV$) can be considered representative of the limit of
infinite heavy-quark masses. Therefore, the $IW$ form factor
$\xi^{(IW)}(\omega)$, obtained through Eq. (\ref{BODY}) using
$w^{T(T_s)}_{(HO)}$, $w^{T(T_s)}_{(conf)}$ and $w^{T(T_s)}_{(GI)}$,
is shown in Fig. 3. It can be seen that in a wide range of values
of the recoil the calculated $IW$ form factor exhibits a moderate
dependence upon the choice of the heavy-meson wave function; in
particular, it is slightly affected by the high-momentum components
present in $w_{(GI)}^{PS}$, as well as by the detailed form of the
confining interaction. Therefore, the $IW$ form factor is found to
be mainly governed by the effects of the confinement scale. Such a
feature can be ascribed to an end-point mechanism, namely to the
fact that the heavy-meson wave function is sharply peaked at $\xi
\sim 1 - m_{sp} / m_Q$, which implies that in Eq. (\ref{BODY})
$|\vec{k'}_{\perp} - \vec{k}_{\perp}| \sim m_{sp} \sqrt{2(\omega -
1)}$, enhancing the contribution due to the low-$k_{\perp}$
($k'_{\perp}$) components of the heavy-meson wave function. It is
only for $\omega > 10 ~ (5)$ that the high-momentum components
present in $w_{(GI)}^{PS}$ can remarkably affect
$\xi^{(IW)}(\omega)$ at $m_{sp} = 0.220 ~ (0.419) ~ GeV$ 
\footnote{Note also from Fig. 3 that, according to the
above-mentioned end-point mechanism, a little bit higher
sensitivity to high-momentum components is found for $\omega < 4$
when $m_{sp} = 0.419 ~ GeV$.}.

\indent Finally, in Fig. 4 our prediction for $\xi^{(IW)}(\omega)$,
calculated at $m_{sp} = 0.220 ~ GeV$, is compared with the results
obtained within various approaches, namely the $ISGW$ prediction
$\xi^{(IW)}(\omega) = e^{- \rho^2 (\omega - 1)}$ with $\rho^2 =
0.60$ \cite{ISGW}, the result of the $RFT$ model \cite{RFT} and the
calculation of Ref. \cite{NEU93}, based on a $QCD$ sum rule
analysis at next-to-leading order in renormalization-group improved
perturbation theory. It can be seen that our result, as well as
the relativistic calculation of Ref. \cite{RFT}, differ only
slightly from the prediction of Ref. \cite{NEU93}, whereas the
$ISGW$ model \cite{ISGW}, characterized by an approximate treatment
of the relativistic effects, predicts higher values for
$\xi^{(IW)}(\omega)$.   

\indent To sum up, the space-like elastic form factor of
heavy-light pseudoscalar mesons has been investigated within a
light-front constituent quark model in order to evaluate the
Isgur-Wise function. It has been shown that in the space-like
sector the Melosh composition of the constituent quark spins plays
a relevant role. Moreover, it turns out that in a wide range of
values of the recoil the calculated Isgur-Wise form factor exhibits
a moderate dependence upon the choice of the heavy-meson wave
function; in particular, $\xi^{(IW)}(\omega)$ is slightly affected
by the high-momentum components generated by the one-gluon-exchange
term of the effective $q \bar{q}$ interaction, as well as by the
detailed form of the confining potential. This fact suggests that
the $IW$ form factor is mainly governed by the effects of the
confinement scale. The slope of $\xi^{(IW)}(\omega)$, calculated at
$m_{sp} = 0.220 ~ (0.419) ~ GeV$, is $\rho^2 = 1.03 ~ (1.14)$, i.e.
it is larger than the $ISGW$ prediction $\rho^2 \sim 0.6 ~ (0.8)$
\cite{ISGW} and in agreement with both $QCD$ sum rule
\cite{BS93,NAR94} and lattice $QCD$ results \cite{LAT1,LAT2}.

\indent The application of our light-front constituent-quark model
to the time-like sector (i.e., in the kinematical regions
pertaining to the weak decays of heavy mesons), as well as the
comparison of the elastic and transition form factors of
heavy-light pseudoscalar mesons in the limit of infinite
heavy-quark masses, are in progress and will be reported elsewhere
\cite{SIM}.   
 
\vspace{1cm}

{\bf Acknowledgment.} The author gratefully acknowledges Fabio
Cardarelli for supplying him with the numerical solutions of Eq.
(\ref{Hqq}).

\newpage

\vspace{0.5cm}

\begin{center}

{\bf Table Captions}

\end{center}

\vspace{0.5cm}

Table 1. The slope $\rho_Q^2 \equiv - [d H_{PS}^Q(\omega) / d
\omega]_{\omega = 1}$ of the body form factor $H_{PS}^Q$ (Eq.
(\ref{BODY}) with $q^2 = 2 M_{PS}^2 (1 - \omega)$), calculated for
the $D$, $B$ and $T$ mesons using different radial wave function
$w^{PS}(k^2)$: $w_{(HO)}^{PS} \propto e^{-k^2 / 2 \beta_{q
\bar{q}}^2}$, $w_{(conf)}^{PS}$ and $w_{(GI)}^{PS}$. The first two
correspond to a quadratic- and linear-confining potential with the
same scale, respectively (see text). The wave function
$w_{(GI)}^{PS}$ is the solution of Eq. (\ref{Hqq}) using for $V_{q
\bar{q}}$ the full $GI$ effective interaction \cite{GI85}. The
result obtained by neglecting the effects of the Melosh rotations
of the $CQ$ spins, i.e. by assuming $R = 1$ in Eq. (\ref{BODY}), is
also reported. The last column corresponds to the asymptotic limit
of infinite heavy-quark masses, calculated through a quadratic
polynomial extrapolation in terms of $1 / M_{PS}$ of the results
obtained at finite values of $M_{PS}$. The mass of the
spectator-quark is $m_{sp} = 0.220 ~ GeV$. For comparison, the
predictions of the $ISGW$ \cite{ISGW} and $RFT$ \cite{RFT} models,
as well as recent results from $QCD$ sum rule ($QCD-SR$)
\cite{BS93,NAR94} and lattice $QCD$ simulations \cite{LAT1}, are
shown.

\vspace{0.5cm}

Table 2. The same as in Table 1, but for the $D_s$, $B_s$ and $T_s$
mesons, using a spectator-quark mass of $0.419 ~ GeV$. For
comparison, the predictions of the $ISGW$ model \cite{ISGW} and the
results of recent lattice $QCD$ calculations \cite{LAT1,LAT2} are
reported.

\newpage

\vspace{0.5cm}

\begin{center}

{\bf Figure Captions}

\end{center}

\vspace{0.5cm}

Fig. 1. (a) The $CQ$ momentum distribution $|w^{PS}(k^2)|^2$
versus the internal momentum $k$. The dot-dashed, dashed and solid
lines correspond to the $D$, $B$ and $T$ mesons, respectively. Thin
and thick lines correspond to $w_{(GI)}^{PS}$ and $w_{(conf)}^{PS}$,
which are solutions of Eq. (\ref{Hqq}) obtained using for $V_{q
\bar{q}}$ the full $GI$ effective interaction \cite{GI85} and
only its linear-confining term, respectively. (b) The $CQ$
momentum distribution in the B-meson. The dotted, dashed and solid
lines correspond to $w^B_{(conf)}$, $w^B_{(HO)} \propto  e^{-k^2 /
(2 \beta_B^2)}$ (with $\beta_B = 0.417 ~ (GeV/c)$) and $w^B_{(GI)}$,
respectively. The dot-dashed line corresponds to the gaussian-like
wave function of Ref. \cite{ISGW}.

\vspace{0.5cm}

Fig. 2. (a) Body form factor $H_{PS}^Q(\omega)$ (Eq. (\ref{BODY})
with $q^2 = 2 M_{PS}^2 (1 - \omega)$) versus $\omega$, calculated
with $w_{(conf)}^{PS}(k^2)$ which corresponds to the solution of
Eq. (\ref{Hqq}) using for $V_{q \bar{q}}$ the linear-confining term
of the $GI$ effective interaction \cite{GI85}. The dotted, dashed
and solid lines correspond to the $D$, $B$ and $T$ mesons,
respectively. The dot-dashed line is the result obtained for the
$T$ meson by neglecting the effects of the Melosh rotations of the
$CQ$ spins, i.e. by assuming $R = 1$ in Eq. (\ref{BODY}). (b) The
same as in (a) but using $w_{(GI)}^{PS}(k^2)$ which corresponds to
the solution of Eq. (\ref{Hqq}) with the full $GI$ effective
interaction.

\vspace{0.5cm}

Fig. 3. The $IW$ form factor $\xi^{(IW)}(\omega)$ versus $\omega$
for a spectator-quark mass of $0.220 ~ GeV$ (a) and $0.419 ~ GeV$
(b). The dashed, dotted and solid lines correspond to the results
obtained using in Eq. (\ref{BODY}) $w_{(HO)}^{T(T_s)}$,
$w_{(conf)}^{T(T_s)}$ and $w_{(GI)}^{T(T_s)}$, respectively. The
dot-dashed line is the result obtained for the $T ~ (T_s)$ meson
neglecting the effects of the Melosh rotations of the $CQ$ spins
(i.e., assuming $R = 1$ in Eq. (\ref{BODY})).  

\vspace{0.5cm}

Fig. 4. Comparison of the $IW$ form factor $\xi^{(IW)}(\omega)$
calculated in various approaches. The solid line is our result
obtained using in Eq. (\ref{BODY}) $w_{(GI)}^{T}(k^2)$ and a
spectator-quark mass of $0.220 ~ GeV$. The dot-dashed line
corresponds to the $ISGW$ prediction $\xi^{(IW)}(\omega) =
e^{- \rho^2 (\omega - 1)}$ with $\rho^2 = 0.60$ \cite{ISGW}. The
dashed line is the result of the $RFT$ model \cite{RFT} and the
dotted line is the prediction of Ref. \cite{NEU93}, based on a
$QCD$ sum rule analysis at next-to-leading order in
renormalization-group improved perturbation theory.

\newpage

\begin{center}

\vspace{1cm}

{\bf TABLE 1}\\

\vspace{0.75cm}

\begin{tabular} {||c ||c |c |c ||c ||}
\cline{1-5} \cline{1-5}
 $w^{PS}(k^2)$&  $D-meson$&  $B-meson$&  $T-meson$& 
 ~~~~ $\infty$ ~~~~
\\ \cline{1-5}
                    $HO$&   1.01&   1.16&   1.25&    1.26
\\ \cline{1-5}
                 $conf.$&   0.94&   1.06&   1.12&    1.13
\\ \cline{1-5}
                    $GI$&   0.68&   0.88&   1.02&    1.03
\\ \cline{1-5}
          $GI ~ (R = 1)$&   0.39&   0.54&   0.69&    0.70
\\ \hline \hline
   $ISGW ~ {\cite{ISGW}}$& ~& ~& ~&    0.5 $\div$ 0.6
\\ \cline{1-5}
     $RFT ~ {\cite{RFT}}$& ~& ~& ~&    0.93 $\pm$ 0.04
\\ \cline{1-5}
 $QCD-SR ~ {\cite{BS93}}$& ~& ~& ~&    0.70 $\pm$ 0.25
\\ \cline{1-5}
$QCD-SR ~ {\cite{NAR94}}$& ~& ~& ~&    1.00 $\pm$ 0.02
\\ \cline{1-5}
  $UKQCD ~ {\cite{LAT1}}$& ~& ~& ~&    $0.9_{-.3 -.2}^{+.2 +.4}$
\\ \cline{1-5}
\end{tabular}

\vspace{5cm}

{\bf TABLE 2}\\

\vspace{0.75cm}

\begin{tabular} {||c ||c |c |c ||c ||}
\cline{1-5}
 $w^{PS}(k^2)$&  $D_s-meson$&  $B_s-meson$& $T_s-meson$&
 ~~~~ $\infty$ ~~~~
\\ \cline{1-5}
                    $HO$&   1.27&   1.40&   1.48&    1.48
\\ \cline{1-5}
                 $conf.$&   1.23&   1.32&   1.37&    1.37
\\ \cline{1-5} \cline{1-5}
                    $GI$&   0.77&   0.98&   1.14&    1.14
\\ \cline{1-5}
          $GI ~ (R = 1)$&   0.43&   0.65&   0.84&    0.84
\\ \hline \hline
  $ISGW ~ {\cite{ISGW}}$& ~& ~& ~&    0.7 $\div$ 0.8
\\ \cline{1-5}
 $UKQCD ~ {\cite{LAT1}}$& ~& ~& ~&    $1.2_{-.2 -.1}^{+.2 +.2}$
\\ \cline{1-5}
   $BSS ~ {\cite{LAT2}}$& ~& ~& ~&    1.24 $\pm$ 0.26 $\pm$ 0.36
\\ \cline{1-5}
\end{tabular}

\end{center}

\newpage

\begin{figure}

\epsfig{file=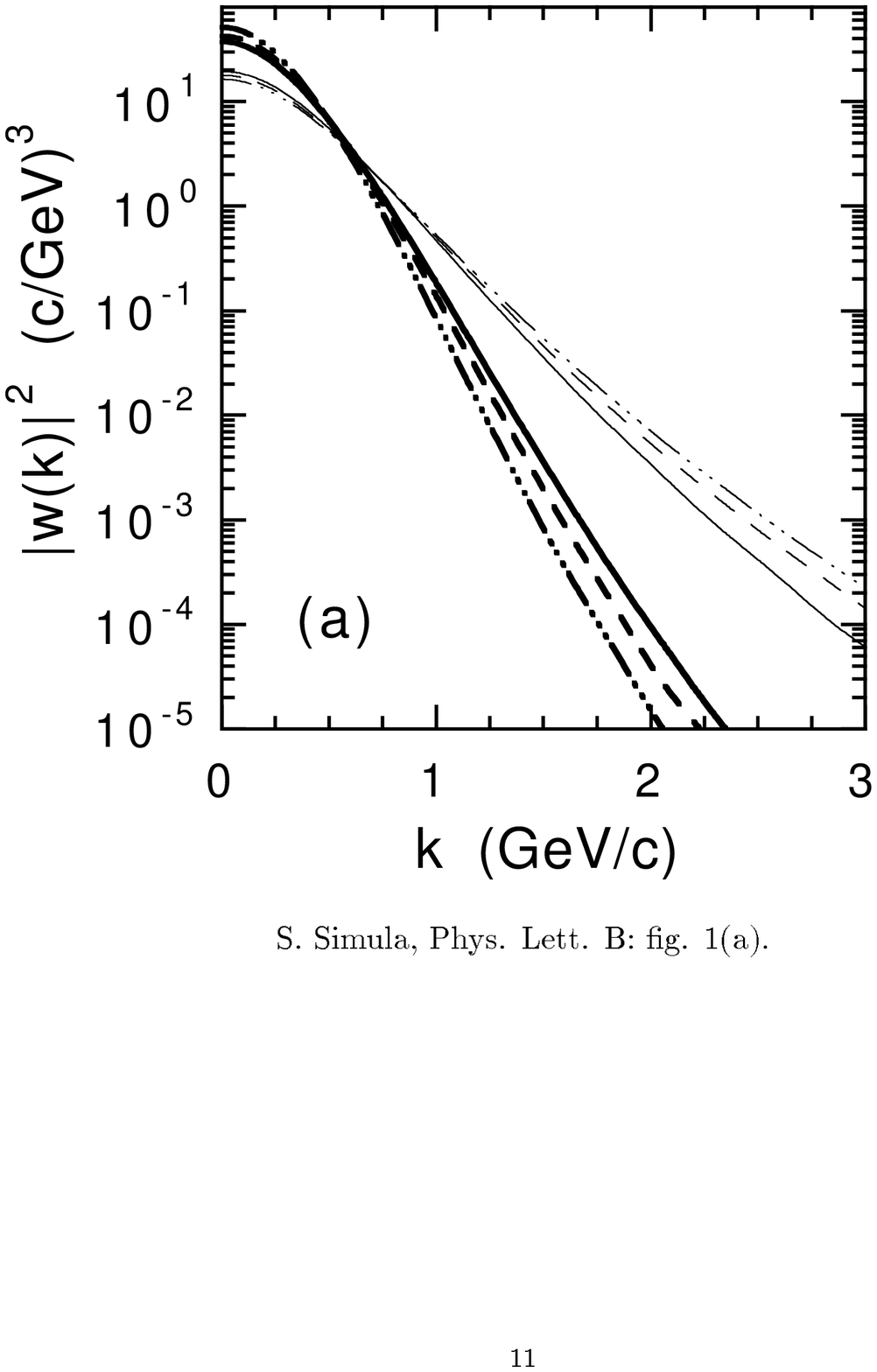}

\end{figure}

\newpage

\begin{figure}

\epsfig{file=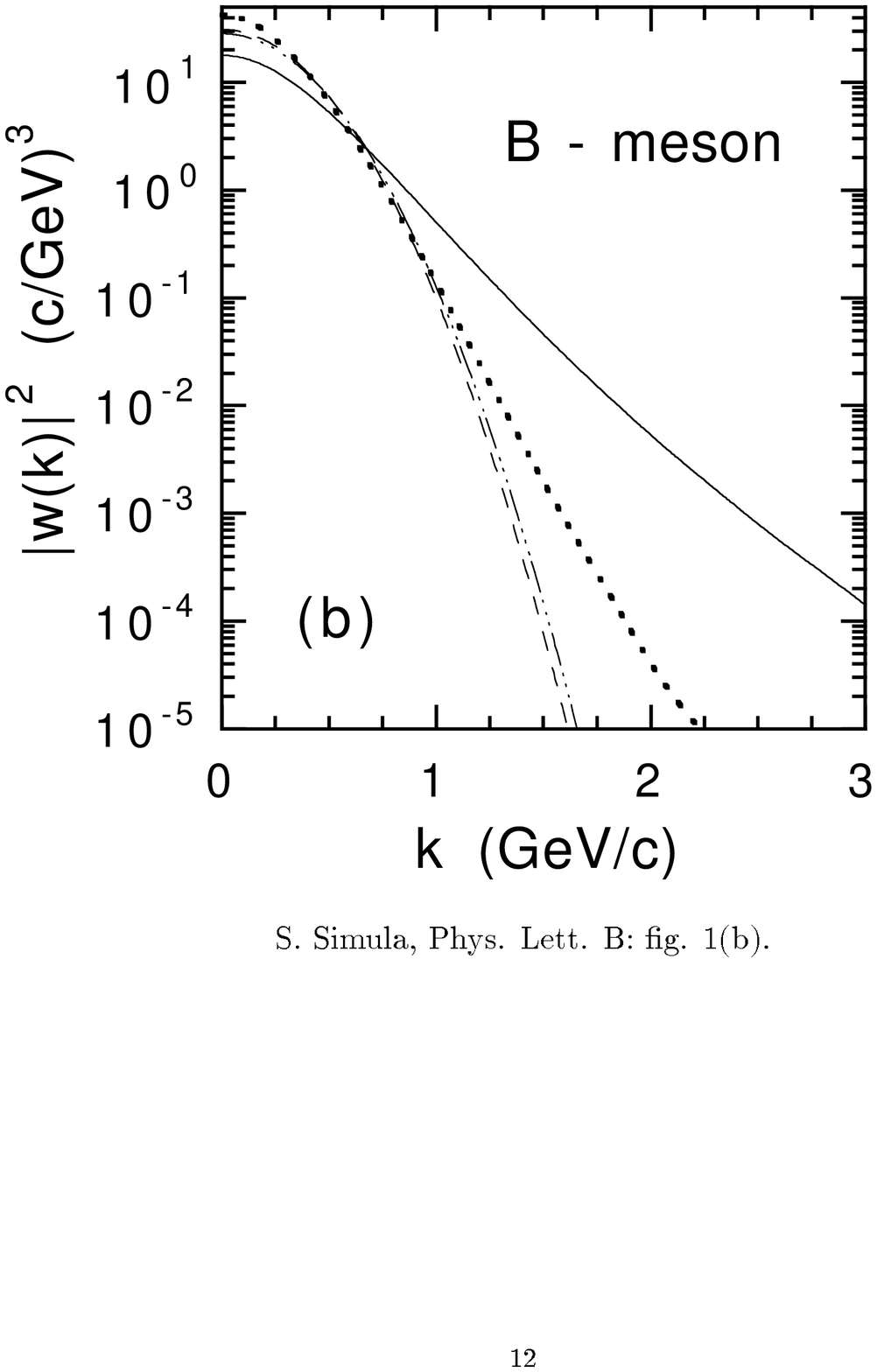}

\end{figure}

\newpage

\begin{figure}

\epsfig{file=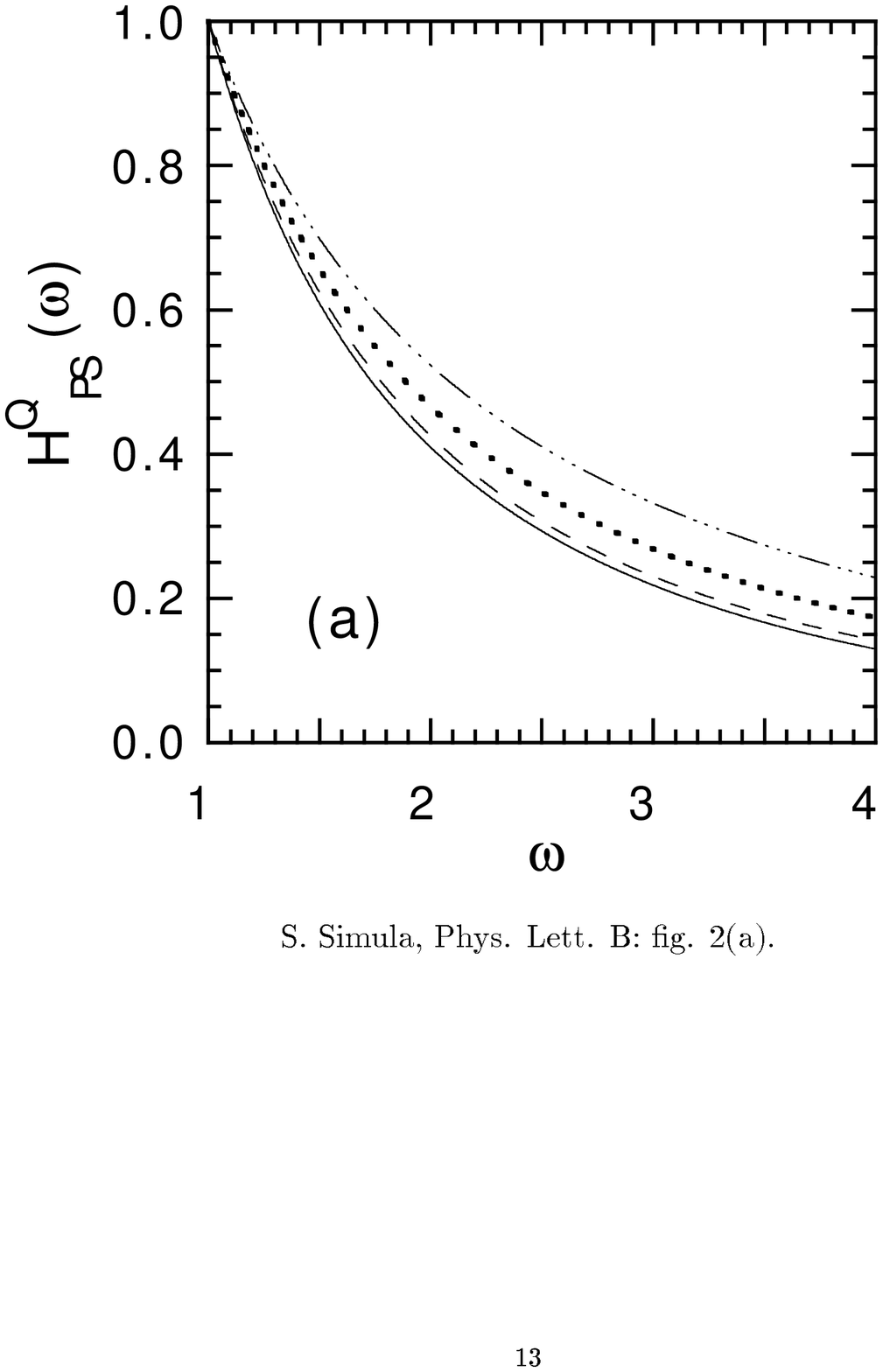}

\end{figure}

\newpage

\begin{figure}

\epsfig{file=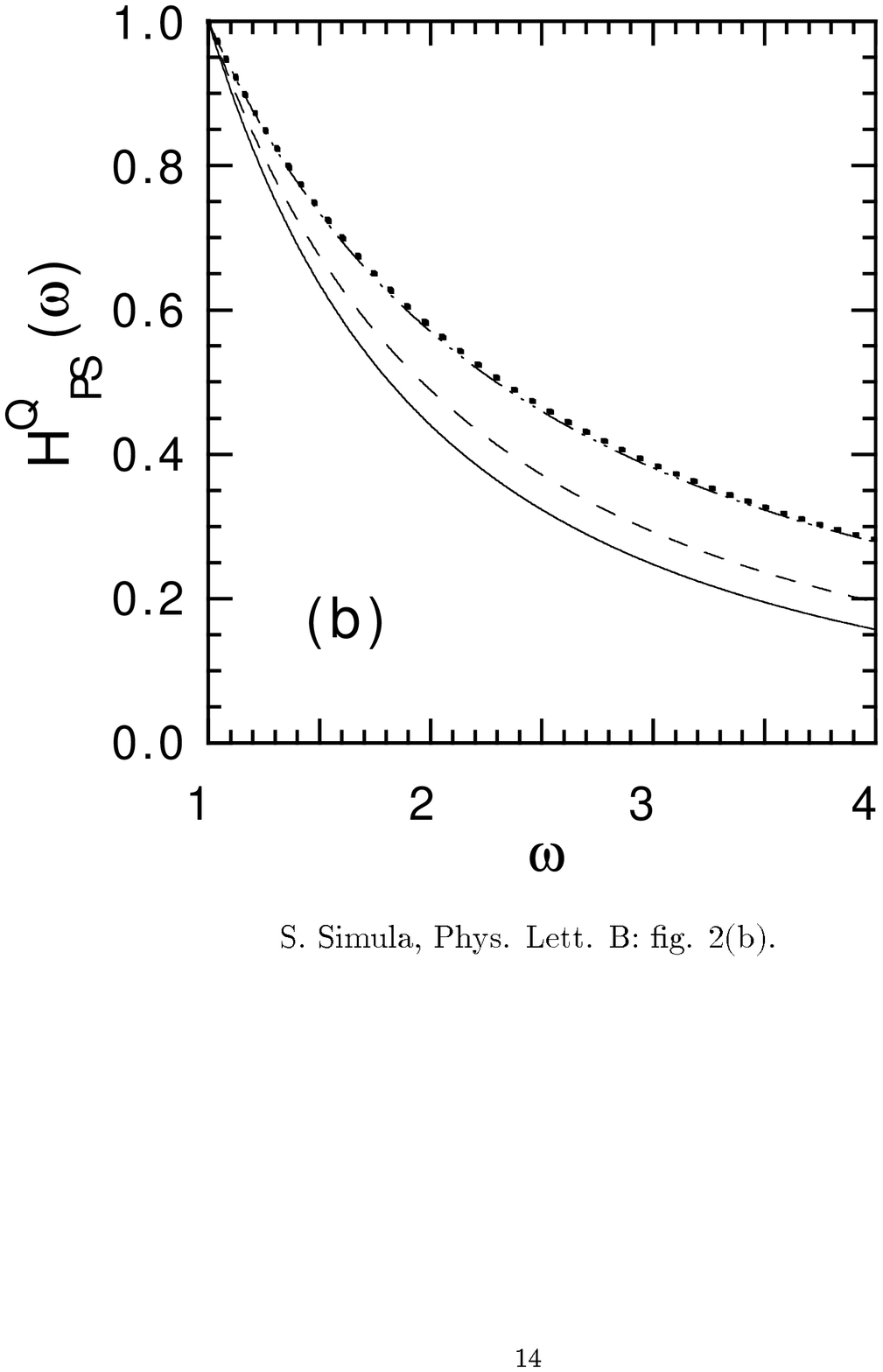}

\end{figure}

\newpage

\begin{figure}

\epsfig{file=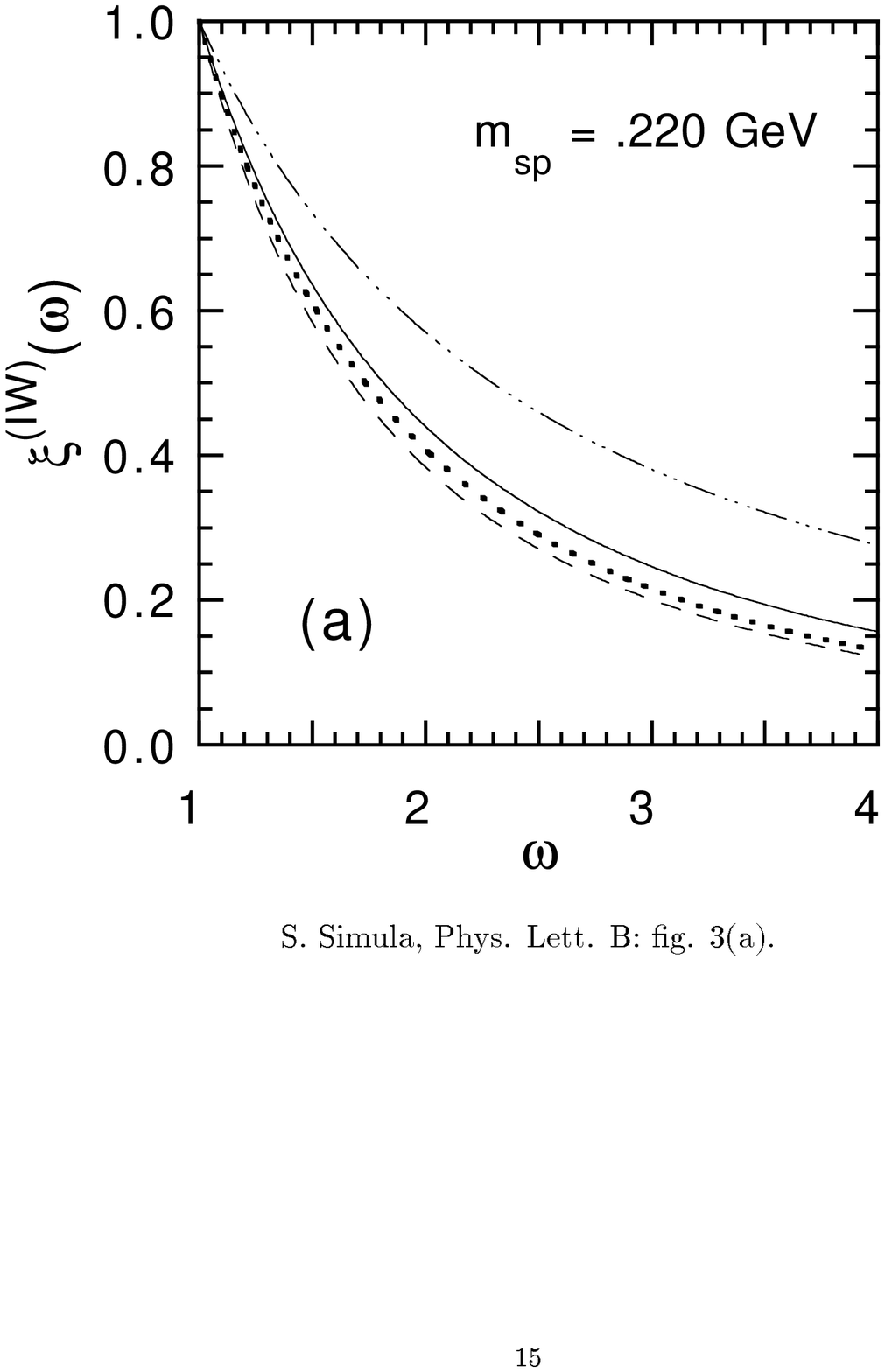}

\end{figure}

\newpage

\begin{figure}

\epsfig{file=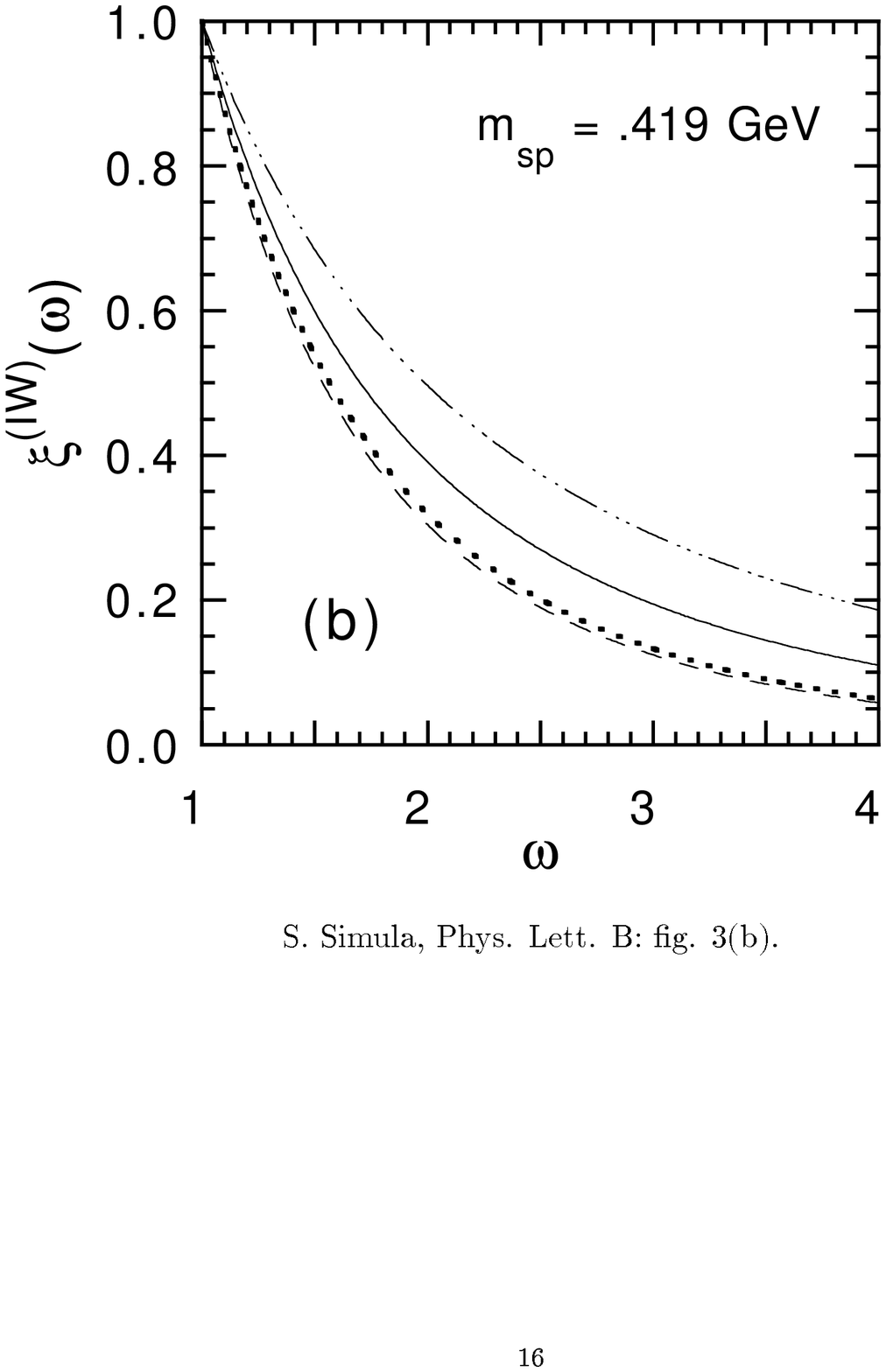}

\end{figure}

\newpage

\begin{figure}

\epsfig{file=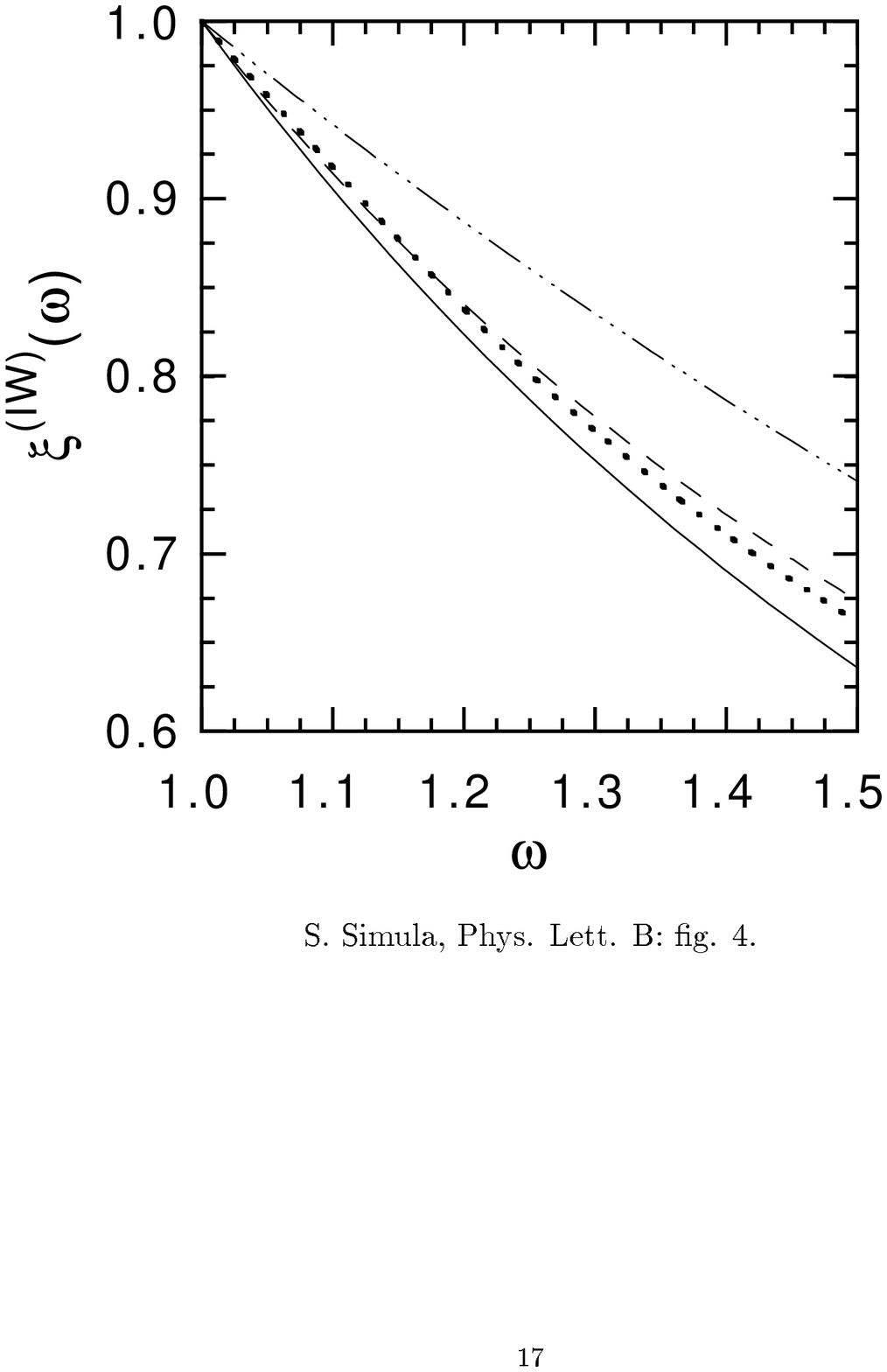}

\end{figure}

\end{document}